\documentclass[runningheads]{llncs}
\usepackage[T1]{fontenc}
\usepackage{graphicx}
\usepackage[hidelinks]{hyperref}
\usepackage{subcaption}
\usepackage{multirow}
\begin{document}
\title{Leaky Batteries: A Novel Set of\\Side-Channel Attacks on Electric Vehicles}
\titlerunning{Leaky Batteries}
\author{Francesco Marchiori\inst{1}\orcidID{0000-0001-5282-0965} \and
Mauro Conti\inst{1}\orcidID{0000-0002-3612-1934}}
\institute{University of Padova, Padova, Italy\\
\email{francesco.marchiori@math.unipd.it, mauro.conti@unipd.it}}
\maketitle              % typeset the header of the contribution
\begin{abstract}
Advancements in battery technology have accelerated the adoption of Electric Vehicles (EVs) due to their environmental benefits.
However, their growing sophistication introduces security and privacy challenges.
Often seen as mere operational data, battery consumption patterns can unintentionally reveal critical information exploitable for malicious purposes.
These risks go beyond privacy, impacting vehicle security and regulatory compliance.
Despite these concerns, current research has largely overlooked the broader implications of battery consumption data exposure.
As EVs integrate further into smart transportation networks, addressing these gaps is crucial to ensure their safety, reliability, and resilience.

In this work, we introduce a novel class of side-channel attacks that exploit EV battery data to extract sensitive user information.
Leveraging only battery consumption patterns, we demonstrate a methodology to accurately identify the EV driver and their driving style, determine the number of occupants, and infer the vehicle's start and end locations when user habits are known.
We utilize several machine learning models and feature extraction techniques to analyze EV power consumption patterns, validating our approach on simulated and real-world datasets collected from actual drivers.
Our attacks achieve an average success rate of 95.4\% across all attack objectives.
Our findings highlight the privacy risks associated with EV battery data, emphasizing the need for stronger protections to safeguard user privacy and vehicle security.

\keywords{Electric Vehicles \and Side-Channel Attacks \and Privacy.}
\end{abstract}
\section{Introduction}
\label{sec:introduction}

Electric Vehicles (EVs) have rapidly transformed the automotive landscape, offering significant environmental benefits and reducing dependence on fossil fuels.
With global EV sales reaching 17.1 million units in 2024, marking a 55\% increase from the previous year, electric mobility is accelerating worldwide~\cite{rhomotion_ev_sales_2024}.
As adoption grows, EVs are becoming increasingly integrated into smart transportation networks, influencing urban mobility, energy consumption, and data-driven vehicle ecosystems.
However, this digitalization raises concerns about data privacy and security, particularly regarding the vast amounts of information generated and transmitted by modern EVs.
In response to these challenges, regulatory frameworks such as the European General Data Protection Regulation (GDPR) and the UNECE WP.29 cybersecurity and software update regulations (R155 and R156) have been established to ensure robust data protection and cybersecurity measures~\cite{unece_wp29_2021,gdpr_2016}.

A critical yet underexplored aspect of EV security is the potential for battery data to leak sensitive information.
Prior research has demonstrated that smartphone lithium-ion batteries can be exploited through side-channel attacks, revealing user behaviors and even cryptographic keys~\cite{la2021wireless,goller2015side}.
Similar vulnerabilities remain largely unexamined in the context of EVs, leaving their full implications uncertain.
Indeed, inference of private details through battery data could enable large-scale surveillance, targeted tracking, or stalking, posing serious privacy risks.
Furthermore, studies have shown that battery data collected during charging can be used to profile EV models, demonstrating that recharging patterns hold identifiable characteristics~\cite{brighente2024evscout2,gangwal2022feasibility}.
Additionally, researchers have highlighted the feasibility of tampering with charging stations, raising concerns about the integrity and confidentiality of data exchanged during the charging process~\cite{conti2022evexchange}.
However, these works primarily focus on recharging capabilities and do not consider battery consumption patterns during driving. As a result, while they provide insights into EV identification based on charging behavior, they do not address how energy usage dynamics during operation can reveal additional distinguishing features.
These factors underscore the urgent need to investigate battery-related privacy threats in EVs.

\paragraph{Contributions.}
In this paper, we introduce a novel set of side-channel attacks designed to extract sensitive user information and uncover confidential EV characteristics.
Our attacks rely solely on battery consumption data and employ time-series feature extraction techniques to train multiple Machine Learning (ML) classifiers.
Our evaluation of these attacks reveals the inherent risks associated with battery consumption patterns and underscores potential attack vectors for exploitation.
Our contributions can be summarized as follows.
\begin{itemize}
    \item We introduce a novel class of side-channel attacks that leverage fine-grained EV battery consumption data analysis to infer sensitive information.
    Unlike previous works that focus primarily on charging patterns, our approach exploits consumption patterns during driving, enabling the identification of previously unexplored privacy threats.
    Specifically, we demonstrate that our attacks can infer the EV driver’s identity and driving style, the vehicle model, the number of occupants, auxiliary power consumption, and the start and end points of a route, particularly when user habits are known.
    To the best of our knowledge, this is the first work to comprehensively analyze these privacy risks from a real-world consumption perspective rather than relying solely on recharging behavior or coarse-grained telemetry.
    \item We define and evaluate several ML classifiers to identify the most effective one for this task. We perform feature extraction from time-series battery consumption data, leveraging the causal nature of the consumption patterns.
    \item We conduct our evaluation using a dataset comprising both simulated and real-world battery consumption traces collected in diverse environments.
    Our results demonstrate an accuracy of up to 0.972 in driving style identification, 0.999 in vehicle identification, 0.907 in occupancy detection, 0.967 in auxiliary power consumption inference, 0.942 in driver identification, 0.955 in trip origin inference, and 0.935 in trip destination inference.
    Therefore, our attacks achieve an average success rate of 95.4\% across all objectives.
    \item We present a proof-of-concept countermeasure that effectively neutralizes our attacks in specific scenarios while significantly reducing their success rate to approximately 45\% in others.
    \item We make our methodology and code open-source at: \texttt{\url{https://anonymous.4open.science/r/Leaky-Batteries-1518/}}.
\end{itemize}

\paragraph{Organization.}
This paper is structured as follows.
Section~\ref{sec:related} reviews related works on EV batteries and associated privacy concerns.
In Section~\ref{sec:systemthreat}, we provide an overview of our system and threat model.
Section~\ref{sec:methodology} details the methodology behind our proposed attacks.
The success of our attacks and the evaluation results are presented in Section~\ref{sec:evaluation}.
We also provide possible countermeasures to our attack in Section~\ref{sec:countermeasures}, followed by our concluding remarks in Section~\ref{sec:conclusions}.
\section{Related Works}
\label{sec:related}

The battery, the central component of EVs, significantly influences the vehicle's performance, efficiency, and lifespan. However, as the reliance on lithium-ion technology grows, so too do concerns about the security and privacy risks associated with it.
Indeed, this growing reliance on lithium-ion batteries has led to concerns about counterfeit products entering the market.
These counterfeit batteries, which pose risks to safety and performance, are increasingly infiltrating the supply chain, with authorities seizing millions of dollars worth annually in various documented cases~\cite{kong2022distribution,obrien2008lithium,us2016cbp}.
Regulatory bodies have introduced stringent safety measures to address these risks, including safety standards for battery construction, usage, and disposal, to ensure that only compliant and safe batteries reach consumers~\cite{berger2022digital}.

In addition to the established safety protocols for EV batteries, data security regulations have become increasingly critical in the context of electric vehicles.
EVs generate vast amounts of sensitive data, such as battery health, driving behavior, and GPS location, which are potential cyberattack targets.
To address these risks, comprehensive cybersecurity regulations have been introduced.
In the European Union, the Cybersecurity Act (EU) 2019/881 mandates that manufacturers implement robust security measures throughout the product lifecycle to prevent data theft and privacy violations~\cite{cybersecurity_act_2019}.
Similarly, the UNECE WP.29 regulation enforces cybersecurity standards and secure software updates for connected and autonomous vehicles~\cite{unece_wp29_2021}.
These regulations extend beyond the vehicles themselves to address vulnerabilities in supporting infrastructure, such as charging stations and cloud platforms, to protect users from privacy breaches and potential cyberattacks.

While regulations aim to protect EV data from direct cyber threats, a growing concern lies in side-channel attacks, which exploit unintended information leakage—such as variations in power consumption, electromagnetic emissions, or timing differences—to infer sensitive data without directly breaking cryptographic systems.
Recent work on smartphones has shown that analyzing power consumption during charging can reveal confidential information~\cite{goller2015side,chawla2019application}.
In battery-powered devices, vulnerabilities in wireless charging interfaces have also been exploited to extract user data~\cite{la2021wireless}.
Extending these insights to the automotive sector, preliminary studies have demonstrated that monitoring charging-current demand can enable the profiling of electric vehicles, potentially compromising user privacy~\cite{brighente2024evscout2}.
Furthermore, research on the feasibility of attacks targeting EV charging infrastructure underscores the emerging risks of data leakage in connected vehicles~\cite{gangwal2022feasibility}.
As EVs become increasingly interconnected, battery consumption data may emerge as a significant side channel for adversaries, underscoring the need for urgent research into these vulnerabilities.
\section{System and Threat Model}
\label{sec:systemthreat}

\begin{figure}
    \centering
    \includegraphics[width=\linewidth]{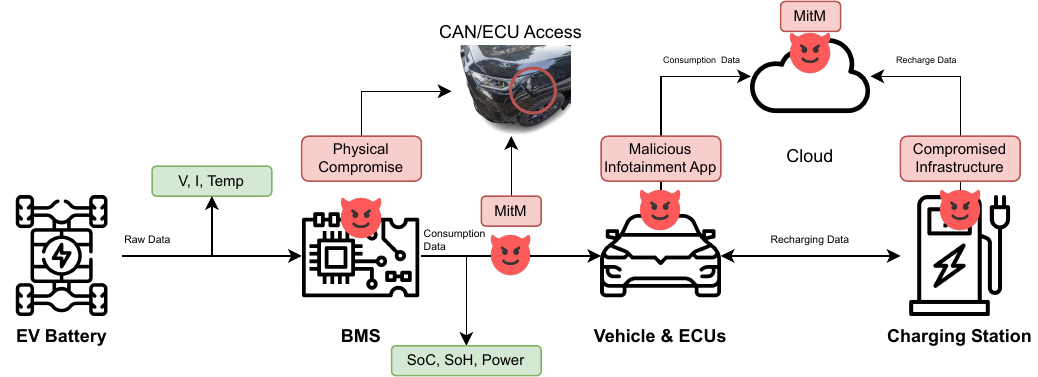}
    \caption{Overview of the system model, the data flow, and the attack vectors. We refer to~\cite{headlights} for CAN/ECU access, and~\cite{conti2022evexchange,brighente2024evscout2} for charging station compromise.}
    \label{fig:framework}
\end{figure}

We present our system and threat model in Section~\ref{subsec:system} and Section~\ref{subsec:threat}.
We showcase the EV battery data collection process, how it is helpful for nominal operations, and how attackers can get a hold of this data and leverage it for malicious purposes.
An overview of the whole model is shown in Fig.~\ref{fig:framework}.

\subsection{System Model}
\label{subsec:system}

EVs generate and process vast amounts of data to ensure efficient operation, safety, and longevity.
At the core of the battery data ecosystem lies the Battery Management System (BMS), a critical component responsible for monitoring and optimizing the performance of lithium-ion batteries~\cite{mishra2021review}.
The BMS continuously collects real-time battery parameters, such as voltage, current, temperature, State of Charge (SoC), and State of Health (SoH) to manage power distribution and prevent potential failures.
This data is transmitted within the vehicle via internal communication protocols, such as the Controller Area Network (CAN), and may also be shared externally through telematics systems for remote diagnostics and fleet management.

Beyond the BMS, modern vehicles incorporate a network of sensors and computing units that monitor driving behavior, energy consumption, and vehicle location.
These systems enable key functionalities such as range estimation, regenerative braking control, and predictive maintenance.
In EVs, battery consumption data is closely linked to these parameters, making it a valuable resource for performance analysis and optimization~\cite{yu2012driving,chung2020optimization,arandia2022analyzing}.
Furthermore, charging session data, including charge rate, energy input, and station location, is logged to enhance battery longevity and improve grid integration.

\subsection{Threat Model}
\label{subsec:threat}

In real-world scenarios, attackers could exploit various aspects to access EV battery consumption data, leveraging different access levels and capabilities.
One potential attack vector involves compromising the BMS or other onboard vehicle sensors through direct physical access~\cite{headlights} or remote cyberattacks, such as exploiting vulnerabilities in connected interfaces (e.g., telematics units, infotainment systems, or diagnostic ports)~\cite{miller2014survey,miller2015remote}.
In this attack scenario, the infotainment system can also be exploited, as Android apps running on Android Auto or Automotive OS can access various vehicle-related data. Consequently, an attacker could develop malicious apps, distribute them through the app store, and covertly collect this data without needing explicit permission access~\cite{pese2023first}.
Additionally, attackers could intercept wireless communications between the vehicle and cloud services, charging stations, or mobile applications, allowing them to collect battery consumption data through passive eavesdropping or active Man-in-the-Middle (MitM) attacks~\cite{conti2022evexchange}.
Another viable approach is data inference through compromised infrastructure, such as tampering with public charging stations to extract charging profiles, which can reveal insights into vehicle usage patterns and driver behaviors~\cite{brighente2024evscout2}.
Even without direct access to the vehicle, an attacker could leverage side-channel analysis, using external sensors or power grid fluctuations to infer EV battery consumption patterns.

Once obtained, this data can be exploited for various malicious purposes.
Indeed, access and consequential inference of private information from this data could reconstruct sensitive behavioral patterns, allowing adversaries to deduce EV users' habits, routines, and even personal preferences~\cite{kang2023electric}.
This could lead to targeted profiling, where individuals are categorized based on their driving style, energy consumption, or mobility patterns, potentially resulting in privacy violations or discriminatory practices.
Furthermore, attackers could manipulate this data to create deceptive user activity models (e.g., usage-based insurance)~\cite{efatinasab2023authentication}, influencing decision-making processes related to insurance, fleet management, or regulatory compliance.
In more advanced scenarios, such insights could be leveraged for coordinated cyber-physical attacks, exploiting energy usage patterns to disrupt vehicle operations or manipulate broader power grid dynamics.
\section{Methodology}
\label{sec:methodology}

This section outlines the methodology and techniques of our proposed side-channel attacks.
We begin by defining the specific objectives of our attacks in Section~\ref{subsec:attacks}, detailing the private information we aim to infer.
Next, Section~\ref{subsec:features} describes our data processing pipeline and feature extraction approach.
Finally, in Section~\ref{subsec:models}, we introduce the machine learning models employed in our attacks and discuss the methodology used for their optimization and tuning.

\subsection{Attacks}
\label{subsec:attacks}

Our side-channel attacks aim to infer sensitive information from EV battery consumption data. Specifically, we target the following aspects.
\begin{itemize}
    \item \textit{Driving Style} --
    We infer driving behaviors such as aggressive, neutral, and defensive driving by analyzing battery consumption patterns.
    These patterns can reveal driver tendencies and habits.
    \item \textit{Vehicle Identification} --
    Variations in energy consumption allow us to distinguish between different EV models based on unique power profiles.
    \item \textit{Occupancy Detection} --
    The energy drawn from an EV battery can be influenced by passenger load, allowing us to estimate the number of occupants in the vehicle.
    \item \textit{Auxiliary System Usage} --
    We infer the power consumption of auxiliary systems (e.g., air conditioning, heating, infotainment), which can provide insights into user comfort preferences and environmental conditions.
    \item \textit{Driver Identification} --
    We identify individual drivers through consistent driving behavior patterns and energy consumption signatures.
    \item \textit{Route Inference} --
    By assuming knowledge of the user's habits, i.e., the locations they frequently visit and their regular driving routes, it becomes possible to infer the following.
    \begin{itemize}
        \item \textit{Origin Identification} --
        We estimate where a trip originates by analyzing initial battery consumption and energy draw patterns.
        \item \textit{Destination Identification} --
        Similarly, consumption patterns allow us to infer the likely endpoint of a trip.
    \end{itemize}
\end{itemize}
These attack objectives highlight the privacy risks associated with battery consumption data, emphasizing the need for stronger protections against inference.

\subsection{Feature Extraction}
\label{subsec:features}

Our analysis relies on battery-related data collected at regular intervals.
The specific features may vary depending on the dataset used (which will be discussed in Section~\ref{subsec:dataset}), but the key features consistent across all our attacks and evaluations are as follows.
\begin{itemize}
    \item Actual Battery Capacity (Wh): the effective energy storage capacity of the battery in watt-hours.
    \item SoC (\%): the percentage of charge remaining.
    \item Total Energy Consumed (Wh): the energy consumed up to a specific step.
    \item Total Energy Regenerated (Wh): the energy recovered via regenerative braking up to a specific step.
    \item Consumption Average (mWh): the consumption average up to a specific step.
\end{itemize}
In addition to these, we also use features such as State of Health (SoH), which indicates the battery's condition and aging; motor power, which tracks the power usage of the motor; torque (Nm), indicating the torque output of the motor; RPM (Revolutions Per Minute), measuring the rotational speed of the motor.

\begin{figure}
    \centering
    \includegraphics[width=\linewidth]{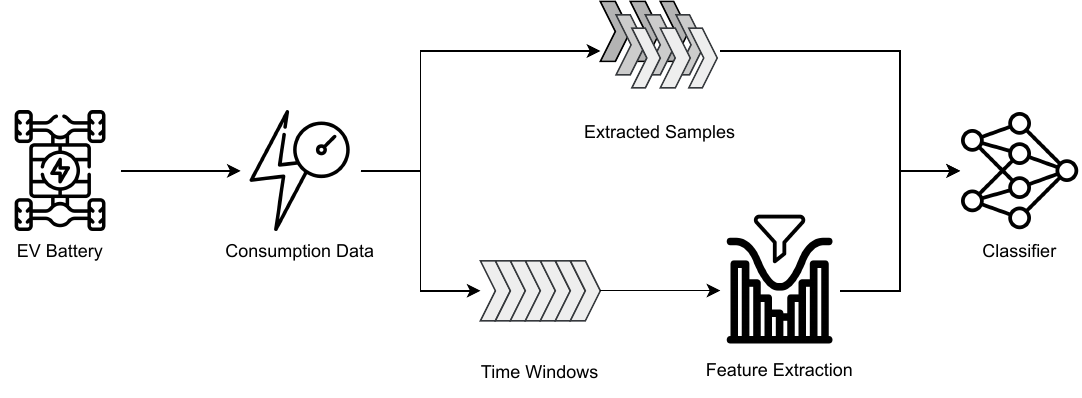}
    \caption{Schema of the feature extraction process. Driving style and vehicle inference attacks operate by simply extracting samples from the consumption data, while every other presented attack leverage the time-series feature extraction pipeline.}
    \label{fig:features}
\end{figure}

This raw data is processed and used by ML classifiers.
An overview of this process is shown in Fig.~\ref{fig:features}.
For driving style and vehicle identification, the correlation between these features is already sufficient for classification at a per-sample level, requiring minimal additional processing.
However, for more complex inferences, further feature engineering is needed.
We leverage \texttt{tsfresh}, an automated Python feature extraction library designed for time-series analysis to extract meaningful insights from sequential data~\cite{christ2018time}.
This library computes a broad range of statistical features, including:
\begin{itemize}
    \item \textit{Basic statistics} – Mean, variance, skewness, and kurtosis.
    \item \textit{Fourier transformations} – Frequency-domain features capturing periodic patterns.
    \item \textit{Temporal characteristics} – Autocorrelation and trend-based metrics.
    \item \textit{Peak and energy-based descriptors} – Maximum values, quantiles, and energy distribution in time series.
\end{itemize}
After extracting the features, each processed dataset is imputed to handle missing or incomplete data effectively.
Once the datasets are imputed, the most relevant features are selected based on the specific objectives of the attack.
To infer occupancy and auxiliary power consumption, we analyze complete driving events (i.e., from trip start to finish) since cumulative energy patterns influence these features.
For driver authentication, we segment data into time windows of 10 samples, as a driver’s control over the vehicle manifests through short-term behavioral patterns.
Finally, for origin and destination inference, we use 5-sample time windows. %, as the critical transition points at the start and end of a trip provide enough discriminative information.
For these types of identification, we focus exclusively on the initial or final percentages of the entire driving event, focusing the analysis on the specific task at hand.
This structured approach ensures that each classification task is supported by an appropriate temporal resolution, balancing granularity and computational efficiency.

\subsection{Models}
\label{subsec:models}

We leverage ML models in this study due to their proven effectiveness in extracting meaningful information from complex battery-related data, as demonstrated in previous works~\cite{marchiori2023your,brighente2024evscout2}.
To assess the performance of each model in the context of our specific tasks, we first perform a grid search to optimize hyperparameters, followed by a train/test split (80\%/20\%) on the \texttt{tsfresh}-processed dataset.
The models employed in our experiments include the following.
\begin{itemize}
    \item \textit{Decision Tree (DT)}~\cite{quinlan1986induction} -- A classification algorithm that recursively splits the data based on feature values to create a tree structure.
    \item \textit{K-Nearest Neighbors (KNN)}~\cite{fix1985discriminatory} -- A simple instance-based learning algorithm that classifies a new point based on the majority class of its nearest neighbors.
    \item \textit{Neural Network (MLP)}~\cite{rumelhart1986learning} -- A model consisting of multiple layers of interconnected nodes (neurons) that can learn complex patterns in the data. 
    \item \textit{Random Forest (RF)}~\cite{breiman2001random} -- An ensemble learning method that constructs a collection of decision trees and aggregates their outputs.
\end{itemize}
Each model undergoes hyperparameter tuning via grid search.
For Decision Trees, we vary the criterion (gini or entropy) and search for the optimal maximum depth within [3, 15].
K-Nearest Neighbors is tested with different values of k in [1, 14] and weighting schemes (uniform or distance).
The Neural Network is evaluated with hidden layer sizes of (50,) and (100,), using ReLU activation and the Adam optimizer.
Finally, Random Forest models are tested with gini as the criterion and the number of estimators set to 100 or 200.
We reduced the hyperparameter search space with respect to prior works involving ML and battery data, as preliminary experiments indicated that excessive fine-tuning had diminishing returns~\cite{marchiori2023your}.
Given the nature of the dataset and the classification tasks, complex optimization was unnecessary to achieve competitive performance.
\section{Evaluation}
\label{sec:evaluation}

We now evaluate the effectiveness of our proposed attacks and analyze their results.
In Section~\ref{subsec:metrics}, we define the evaluation metrics relevant to our ML tasks.
Next, in Section~\ref{subsec:dataset}, we provide an overview of the datasets used in our experiments, detailing their characteristics.
In Section~\ref{subsec:results}, we present the success rates of our attacks.
In Section~\ref{subsec:importance}, we provide an analysis of feature importance for our classifiers, followed by a discussion of the attack's implications in Section~\ref{subsec:discussion}.

\subsection{Metrics}
\label{subsec:metrics}

Since our attacks rely on ML models to classify various attributes (such as driving style, vehicle type, occupancy, and origin/destination), the success of these attacks is tied directly to the performance of the models.
From a ML perspective, our task is primarily a multiclass classification problem.
For example, the number of occupants is limited to a range from 1 to 5, and we have predefined categories for driving styles.
Additionally, we assume knowledge of user habits for the classification of origin and destination, enabling us to work with a fixed set of labels.

For each classification task, we use two standard performance metrics: accuracy and F1 score.
Although it is possible to define the accuracy for each label in terms of True Positives (TPs), False Positives (FP), False Negatives (FNs), and True Negatives (TNs), for our evaluation, we will consider overall accuracy defined as follows.
\begin{equation}
    \mathrm{Accuracy} = \frac{\mathrm{Number\:of\:Correct\:Predictions}}{\mathrm{\:Total\:Predictions}}.
\end{equation}
Since accuracy directly reflects the model's performance in distinguishing targets, it effectively represents the attack's success rate.
We also compute the macro-average F1 score, which provides a balanced assessment of precision and recall across all classes. While this metric is particularly useful for handling class imbalances, the dataset we use maintains a relatively even distribution of $N$ labels across the evaluated tasks.
\begin{equation}
    F1_{macro} = \frac{1}{N} \sum\limits^N_{i=1} F1_i = \frac{1}{N} \sum\limits^N_{i=1} 2 \frac{\mathrm{Precision}_i \cdot \mathrm{Recall}_i}{\mathrm{Precision}_i + \mathrm{Recall}_i};
\end{equation}
where
\begin{equation}
    \mathrm{Precision}_i = \frac{TP_i}{TP_i+FP_i},
\end{equation}
\begin{equation}
    \mathrm{Recall}_i = \frac{TP_i}{TP_i+FN_i}.
\end{equation}

\subsection{Dataset}
\label{subsec:dataset}

This study leverages a dual EV dataset developed to enhance battery range estimation using ML models trained on simulated and real-world driving data.
Given our ML-based approach to analyzing battery-related characteristics, this dataset is particularly well-suited to our research.
The dataset originates from a study that employs the Simulation of Urban MObility (SUMO) software to generate synthetic data and also collects real-time data from electric vehicles using onboard diagnostics (OBD-II) and external APIs for environmental conditions~\cite{cejudo2024electric}.
The dataset enables an in-depth analysis of energy consumption factors, providing insights beyond standard manufacturer consumption values.

\begin{figure}[!htbp]
    \centering
    \begin{subfigure}{0.495\textwidth}
        \centering
        \includegraphics[width=\textwidth]{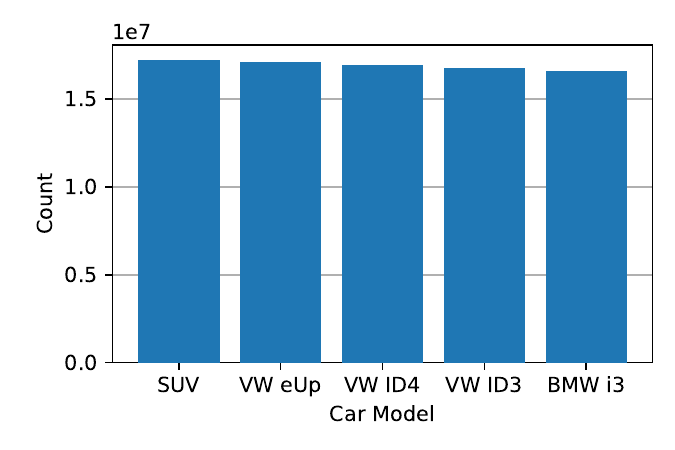}
        \caption{Car Model.}
        \label{subfig:car_model}
    \end{subfigure}
    \begin{subfigure}{0.495\textwidth}
        \centering
        \includegraphics[width=\textwidth]{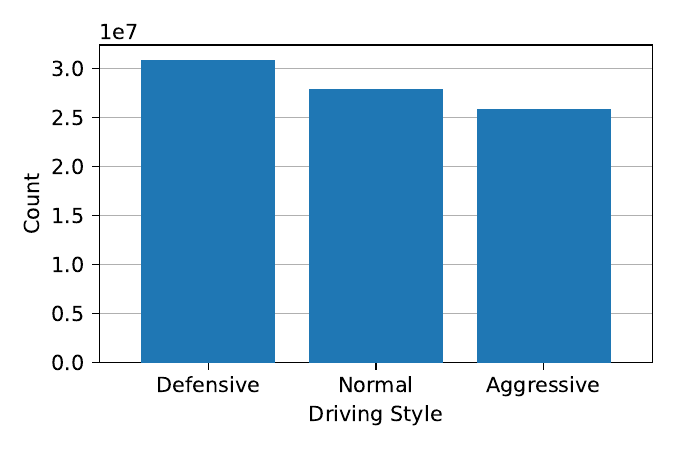}
        \caption{Driving Style.}
        \label{subfig:driving_style}
    \end{subfigure}
    \begin{subfigure}{0.495\textwidth}
        \centering
        \includegraphics[width=\textwidth]{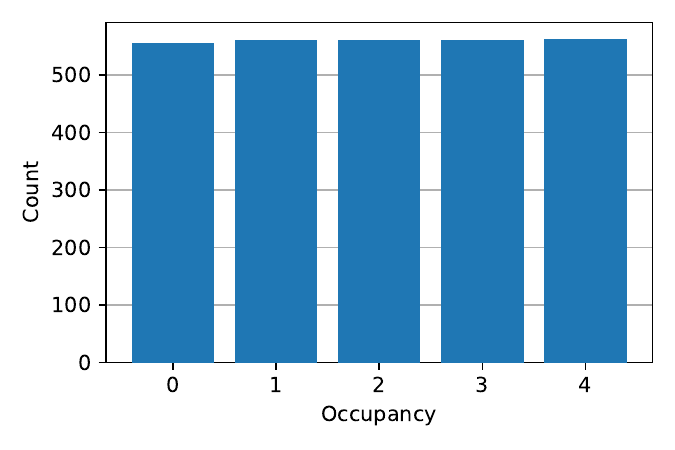}
        \caption{Occupancy.}
        \label{subfig:occupancy}
    \end{subfigure}
    \begin{subfigure}{0.495\textwidth}
        \centering
        \includegraphics[width=\textwidth]{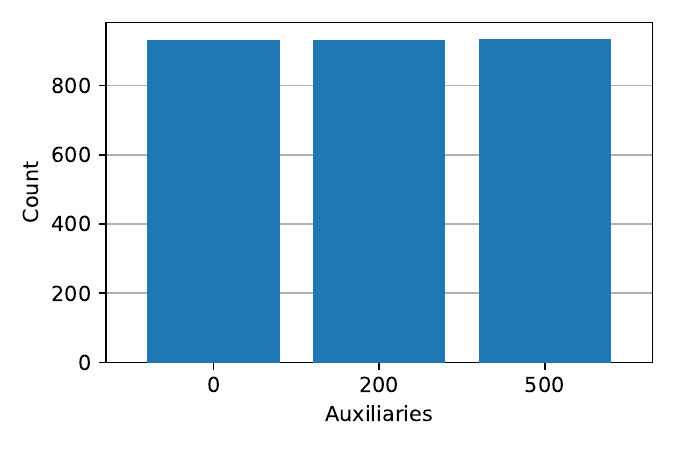}
        \caption{Auxiliary consumption (Watts).}
        \label{subfig:auxiliary}
    \end{subfigure}
    \caption{Statistics for attack targets in the simulated dataset.}
    \label{fig:simulated}
\end{figure}

\paragraph{Simulated Dataset.}
The Dataset of Electric Vehicle Synthetic Trips (DEVST) consists of electric vehicle trip data generated using SUMO, covering 21 routes with a wide range of driving conditions.
Each route is simulated by varying multiple parameters, including driver behavior (aggressive, moderate, defensive), vehicle occupancy (1 to 5 passengers), auxiliary power consumption, wind conditions, and traffic levels.
This combination results in a total of 42,525 trip simulations.
Each simulation records parameters such as vehicle speed, SoC, total energy consumed, road slope, and remaining range.
Multiple electric vehicle models, including BMW i3, VW ID.3, VW ID.4, VW e-Up, and a generic SUV, are simulated under different driving styles.
For this dataset, we perform side-channel attacks to infer driving style, vehicle type, occupancy level, and auxiliary power consumption.
We present data balance of these metrics in Fig.~\ref{fig:simulated}.

\begin{figure}[!htbp]
    \centering
    \begin{subfigure}{0.5\textwidth}
        \centering
        \includegraphics[width=\textwidth]{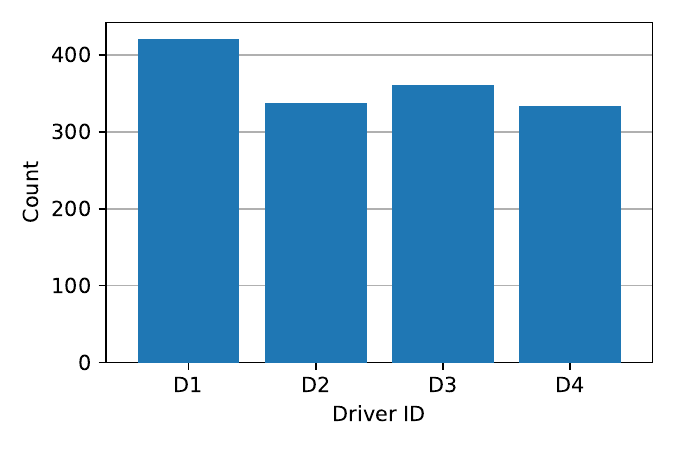}
        \caption{Drivers.}
        \label{subfig:drivers}
    \end{subfigure}
    \begin{subfigure}{0.495\textwidth}
        \centering
        \includegraphics[width=\textwidth]{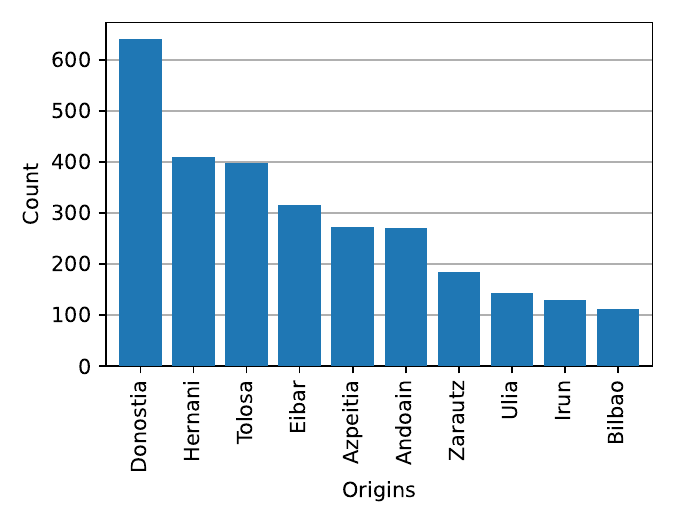}
        \caption{Trip origin.}
        \label{subfig:origins}
    \end{subfigure}
    \begin{subfigure}{0.495\textwidth}
        \centering
        \includegraphics[width=\textwidth]{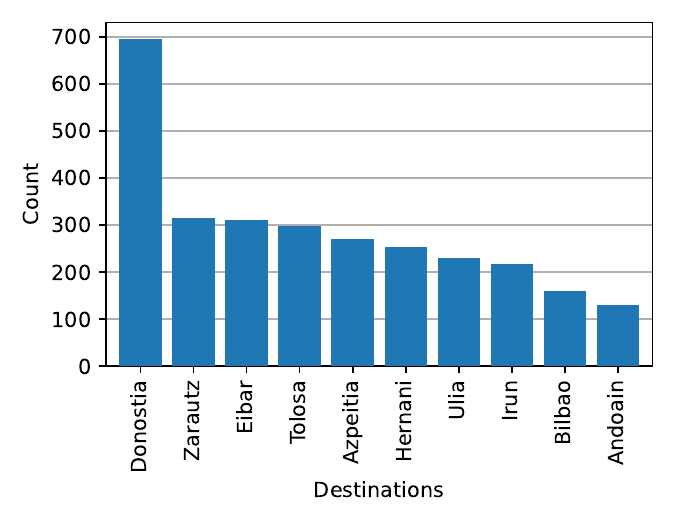}
        \caption{Trip Destination.}
        \label{subfig:destinations}
    \end{subfigure}
    \caption{Statistics for attack targets in the real dataset. Counts shown here already take into account the time-window processing overviewed in Section~\ref{subsec:features}.}
    \label{fig:real}
\end{figure}

\paragraph{Real-World Dataset.}
The Dataset of Electric Vehicle Real Trips (DEVRT) consists of 58 trips recorded using two electric vehicles: a Nissan Leaf e+ 62kWh and a Dacia Spring 33kW.
These trips, carried out over four consecutive days in the Basque Country, Spain, include both urban and medium-distance routes with varying elevation profiles.
Data collection involved an OBD-II device for vehicle telemetry and external APIs for traffic and weather conditions.
The dataset captures data types such as battery state of charge, speed, power consumption, energy regeneration, environmental conditions (temperature, wind speed/direction), and real-time road traffic data.
For this dataset, we conduct side-channel attacks to infer the driver's identity, trip origin, and trip destination.
We present the data distribution of these metrics in Fig.~\ref{fig:real}.
Since the distribution of origin and destination classes is highly imbalanced, we perform experiments on both the original dataset and a randomly undersampled version to achieve class balance.
The original dataset reflects a realistic scenario where a user may frequently visit specific locations (e.g., home or work) while traveling to others less often.
In contrast, the undersampled dataset provides a more balanced distribution, aligning better with the assumptions of typical machine learning tasks.

\subsection{Attacks Results}
\label{subsec:results}

We now show the results of our experiments and the effectiveness of our attack.
Notably, given the specific metrics defined in Section~\ref{subsec:metrics}, accuracy in this context directly corresponds to the attack success rate when evaluating the effectiveness of the ML models.
It is also worth noting that although our results present different attack objectives separately, they can be inferred simultaneously as long as the required features are available, i.e., EV battery consumption data. 

\begin{figure}[!htbp]
    \centering
    \begin{subfigure}{0.495\textwidth}
        \centering
        \includegraphics[width=\textwidth]{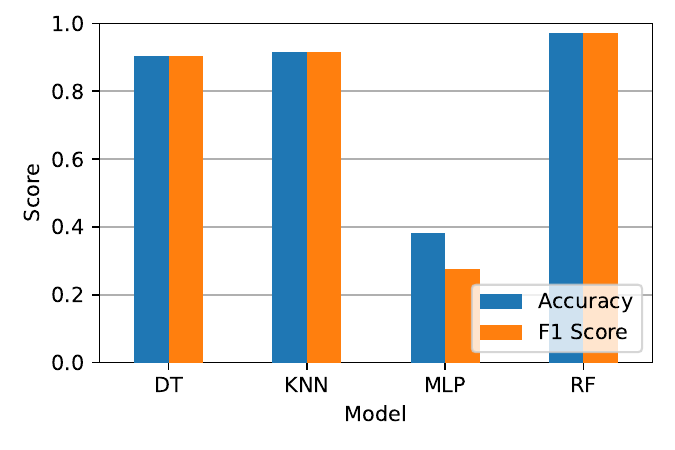}
        \caption{Driving style inference.}
        \label{subfig:style_results}
    \end{subfigure}
    \begin{subfigure}{0.495\textwidth}
        \centering
        \includegraphics[width=\textwidth]{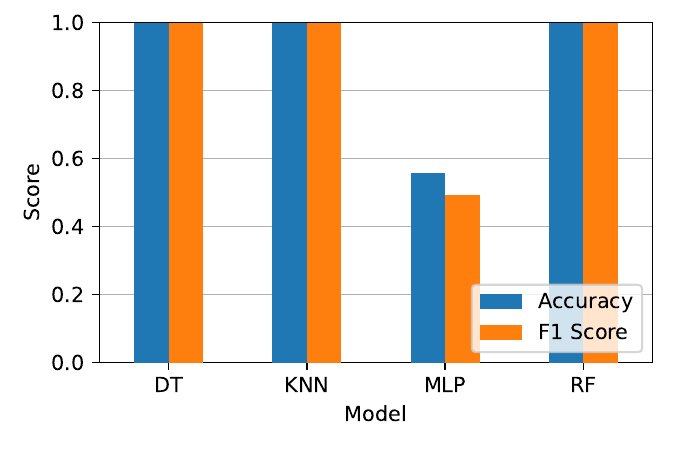}
        \caption{Car model inference.}
        \label{subfig:car_results}
    \end{subfigure}
    \caption{Results for attacks in the simulated dataset leveraging single sample feature correlation.}
    \label{fig:simulated_results1}
\end{figure}

\paragraph{Simulated Dataset.}
As anticipated in Section~\ref{subsec:features}, driving style and vehicle identification attacks do not require time-series analysis of the battery consumption data.
As such, we perform inference at each data sample of the simulated dataset, obtaining the results shown in Fig.~\ref{fig:simulated_results1}.
As we can see from the graphs, our models can infer each attack objective with almost perfect scores.
The best results are obtained in car model inference, which is expected given the strong correlation between battery characteristics and vehicle type.
Previous research has already demonstrated the feasibility of battery authentication, and since our features are closely related to those used in such studies, our results reinforce that battery consumption patterns can serve as a reliable identifier~\cite{marchiori2023your}.
However, while prior work leveraged this for legitimate authentication, our study highlights its potential exploitation for adversarial purposes.

\begin{figure}[!htbp]
    \centering
    \begin{subfigure}{0.495\textwidth}
        \centering
        \includegraphics[width=\textwidth]{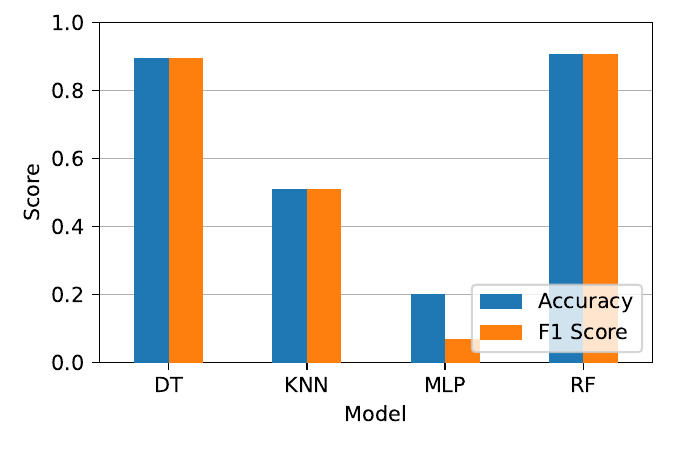}
        \caption{Occupancy inference.}
        \label{subfig:occupancy_results}
    \end{subfigure}
    \begin{subfigure}{0.495\textwidth}
        \centering
        \includegraphics[width=\textwidth]{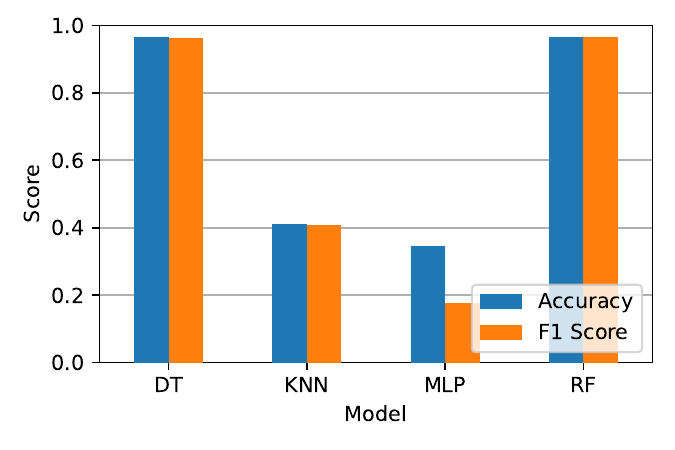}
        \caption{Auxiliary consumption inference.}
        \label{subfig:auxiliaries_results}
    \end{subfigure}
    \caption{Results for attacks in the simulated dataset leveraging single sample feature correlation.}
    \label{fig:simulated_results2}
\end{figure}

In Fig.~\ref{fig:simulated_results2}, we show the results for the time series-related attack objectives, i.e., occupancy and auxiliary power consumption.
The evaluation still presents high scores, with a maximum accuracy of 0.907 for the former task and 0.967 for the latter, obtained with the RF model.
These results represent average performance across all trips.
In reality, specific trips can exhibit significantly better inference accuracy.
For instance, occupancy inference shows a standard deviation of 0.092, indicating that some trips are more susceptible to accurate classification than others.
However, since the time-series feature extraction is performed at the trip level, multiple data collection instances on the same trip would be required to validate this hypothesis further.

\paragraph{Real-World Dataset.}

Table~\ref{tab:driver_results} presents the evaluation results for our driver identification attack.
The RF model demonstrates the highest performance among all tested models, which is consistent with previous findings.
However, the absence of labeled driving style data in the real-world dataset limits our ability to explore potential correlations between these results and those depicted in Fig.~\ref{subfig:style_results}, where driving style was analyzed in the synthetic dataset.

\begin{table}[!htpb]
  \centering
  \caption{Results for the driver identification attack.}
  \label{tab:driver_results}
  \begin{tabular}{l|c|c}
    \hline
    \multirow{2}{*}{\textbf{Model\:}} & \multicolumn{2}{c}{\textbf{Scores}} \\ \cline{2-3}
    & \:Accuracy\: & F1 \\
    \hline
    DT & 0.933 & \:0.933\: \\
    KNN & 0.375 & 0.381 \\
    MLP & 0.317 & 0.248 \\
    RF & 0.942 & 0.942 \\
    \hline
  \end{tabular}
\end{table}

Table~\ref{tab:trip_results} presents the evaluation results for the trip inference attacks targeting the prediction of origin and destination cities.
Our models accurately identify these points within a scenario including 10 different cities.
Notably, dataset balancing has minimal impact on performance, suggesting that the models are resilient to class imbalance and can effectively extract distinguishing patterns regardless of the frequency distribution of trip origins and destinations.
This implies that the attack remains highly effective even in real-world settings, where certain locations are visited more frequently than others.

\begin{table}[!htpb]
  \centering
  \caption{Results for the trip inference attack.}
  \label{tab:trip_results}

  \begin{subtable}[t]{0.75\textwidth}
    \centering
    \caption{Origin identification.}
    \label{subtab:origin_results}
    \begin{tabular}{l|c|c|c|c}
      \hline
      \multirow{2}{*}{\textbf{Model\:}} & \multicolumn{2}{c|}{\textbf{\:Balanced Scores\:}} & \multicolumn{2}{c}{\textbf{\:Unbalanced Scores\:}} \\ \cline{2-5}
      & \:Accuracy\: & \:F1\: & \:Accuracy\: & \:F1\: \\
      \hline
      DT  & 0.909 & 0.908 & 0.967 & 0.966 \\
      KNN & 0.182 & 0.180 & 0.279 & 0.275 \\
      MLP & 0.045 & 0.009 & 0.197 & 0.158 \\
      RF  & 0.955 & 0.953 & 0.951 & 0.951 \\
      \hline
    \end{tabular}
  \end{subtable}
  \hfill
  \begin{subtable}[t]{0.75\textwidth}
    \centering
    \caption{Destination identification.}
    \label{subtab:destination_results}
    \begin{tabular}{l|c|c|c|c}
      \hline
      \multirow{2}{*}{\textbf{Model\:}} & \multicolumn{2}{c|}{\textbf{\:Balanced Scores\:}} & \multicolumn{2}{c}{\textbf{\:Unbalanced Scores\:}} \\ \cline{2-5}
      & \:Accuracy\: & \:F1\: & \:Accuracy\: & \:F1\: \\
      \hline
      DT  & 0.885 & 0.877 & 0.952 & 0.948 \\
      KNN & 0.462 & 0.454 & 0.435 & 0.451 \\
      MLP & 0.231 & 0.157 & 0.161 & 0.120 \\
      RF  & 0.923 & 0.927 & 0.935 & 0.936 \\
      \hline
    \end{tabular}
  \end{subtable}
\end{table}

\subsection{Feature Analysis}
\label{subsec:importance}

\begin{figure}
    \centering
    \includegraphics[width=.6\linewidth]{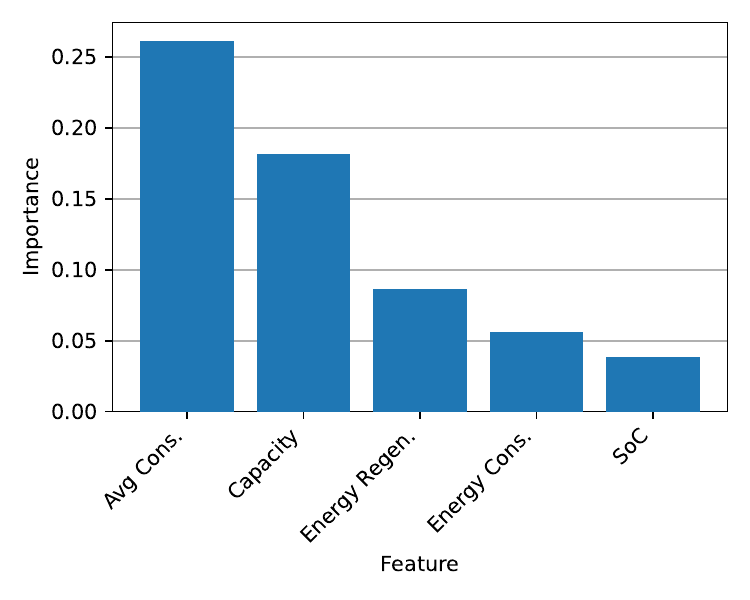}
    \caption{SHAP output for driving style identification with the DT model.}
    \label{fig:shap}
\end{figure}

To better grasp the role of battery consumption data in the classifications performed during our attacks, we employ eXplainable AI (XAI) techniques.
Specifically, we analyze feature importance using SHapley Additive exPlanations (SHAP), a model-agnostic XAI approach~\cite{NIPS2017_7062}.
Due to the linear increase in feature count caused by the time-series processing performed by \texttt{tsfresh}, we focus our analysis on the driving style inference attack, which directly utilizes the battery consumption data without any preprocessing (i.e., the upper flow in Fig.\ref{fig:features}).
The results of this analysis are presented in Fig.\ref{fig:shap}.
The average consumption is the most important feature, closely followed by battery capacity.
In contrast, the SoC feature is the least important, a result that aligns with expectations due to its time-dependent nature.
Since the classifiers process features independently of their causal relationships, the SoC feature has less influence on the inference.

\subsection{Discussion}
\label{subsec:discussion}

Throughout our evaluation, the RF model emerged as the most effective solution across all attack objectives.
Its robustness can be attributed to its ensemble learning approach, which mitigates overfitting by averaging predictions from multiple decision trees.
This dominance of RF aligns with findings in the literature, where similar architectures have shown strong performance in battery data classification tasks~\cite{marchiori2023your}.
In contrast, the MLP model underperformed, yielding notably lower accuracy compared to other works in the field.
This subpar performance can be attributed to the limited search space during the grid search, a deliberate constraint implemented to minimize computational overhead.
While it is true that, in some attack scenarios, data processing and inference could be performed online, our primary goal was to ensure minimal overhead while demonstrating the effectiveness of even simpler ML models for such tasks.
\section{Countermeasures}
\label{sec:countermeasures}

Our attacks successfully infer sensitive vehicle information from the dataset, highlighting a significant security risk.
However, the type of data being exploited remains essential for the nominal operation of vehicles, meaning that removing it entirely is not a viable option.
At the same time, traditional encryption and authentication techniques are challenging to implement on the CAN bus due to its strict bandwidth limitations and legacy design~\cite{lotto2024survey}.
Many existing security solutions require substantial modifications to the original protocol or the adoption of CAN-FD or CAN+, which are not always feasible for deployment in current automotive systems.

\begin{figure}[!htbp]
    \centering
    \begin{subfigure}{0.495\textwidth}
        \centering
        \includegraphics[width=\textwidth]{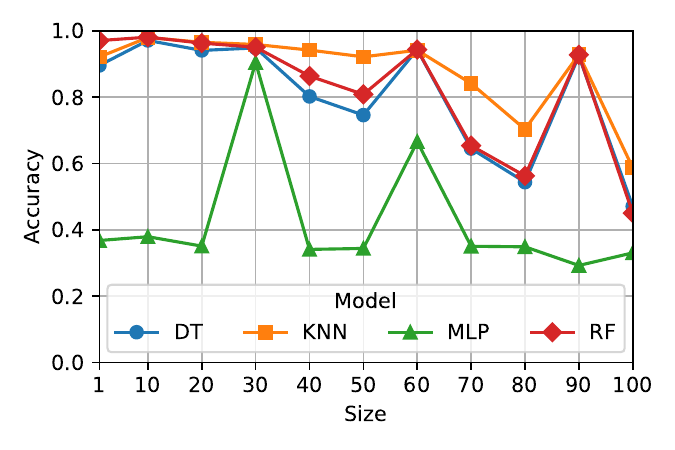}
        \caption{Driving style inference.}
        \label{subfig:countermeasure_style}
    \end{subfigure}
    \begin{subfigure}{0.495\textwidth}
        \centering
        \includegraphics[width=\textwidth]{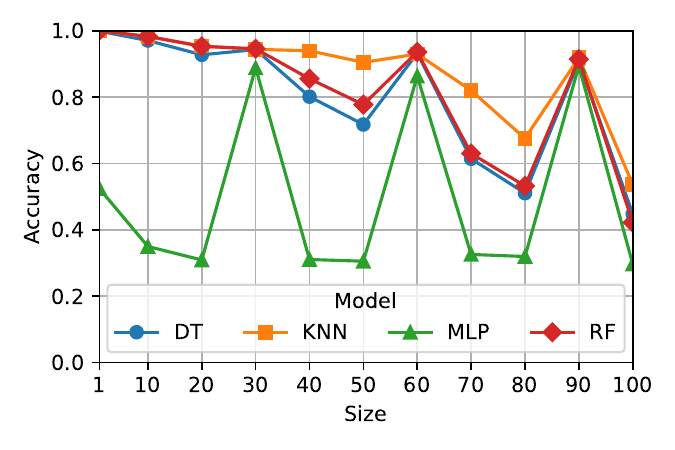}
        \caption{Car model inference.}
        \label{subfig:countermeasure_car}
    \end{subfigure}
    \caption{Countermeasure evaluation for different time window sizes.}
    \label{fig:countermeasure}
\end{figure}

We propose a time-based aggregation technique to address this challenge as a lightweight and effective countermeasure.
By aggregating data over varying time windows, we disrupt the time-series statistical patterns previously exploited by our attacks.
Specifically, many of the successful attacks relied on time-series analysis, computing multiple statistical features over sequential data points.
By aggregating data into windows and computing only the mean over different time intervals, we eliminate the fine-grained temporal relationships necessary for these attacks, making techniques such as using \texttt{tsfresh} ineffective.
However, it is crucial to evaluate the trade-off between information preservation and attack resilience at different window sizes for attacks that do not rely on time-series characteristics, such as driving style and vehicle inference.
To this end, we conduct an in-depth analysis using multiple time window sizes at intervals of 10 samples per window until 100.
The intuition behind this approach is that smaller windows retain more granular vehicle behavior details, benefiting legitimate vehicle functions but providing more information to attackers.
Conversely, larger windows obscure detailed patterns, offering stronger privacy protection at the potential cost of reduced utility.
To systematically evaluate this trade-off, we perform a stratified reduction on the dataset and analyze the accuracy of attacker models at each window size.
The results, summarized in Fig.~\ref{fig:countermeasure}, illustrate the impact of aggregation on attack effectiveness, highlighting the degradation of accuracy (i.e., the success rate of our attacks) up to 0.450 for driving style inference and 0.422 for car model inference (RF accuracy at size = 100). 
However, we observe certain peaks at specific window sizes, which appear as outliers.
When excluded, the overall trend follows a descending pattern.
Therefore, if this countermeasure is implemented, it will be crucial to assess the trade-off between privacy and data utility and carefully select a window size that ensures resilience against such outliers.

\section{Conclusions}
\label{sec:conclusions}

As the adoption of EVs grows, so too does the volume of sensitive data they generate.
Among these data, battery consumption patterns stand out as a potential security and privacy risk.
Despite prior research on battery side-channel attacks in other domains, the implications for EVs remain underexplored.
This paper aimed to address these concerns by investigating the potential for extracting sensitive user information through battery data in the context of electric vehicles.

In this work, we introduced a set of novel side-channel attacks that exploit EV battery consumption data to infer private user information and identify key characteristics of the vehicle.
Through the use of time-series feature extraction and ML classifiers, we demonstrated the feasibility of inferring sensitive details, such as driver identification, trip characteristics, and the number of occupants in the vehicle.
Our comprehensive evaluation using both simulated and real-world datasets highlighted the potential for battery consumption data to be used as an attack vector, achieving significant classification accuracy.

\paragraph{Limitations and Future Works.}
While our study provides important insights into the security risks associated with EV battery data, it is essential to recognize the potential limitations of our approach.
The attacks proposed in this paper assume the availability of battery consumption data, and while this is a realistic scenario in many contexts, the effectiveness of these attacks may vary depending on the noise and variability present in real-world data.
Additionally, our evaluation was conducted on a dataset that, although diverse, may not fully encompass the broad range of driving conditions, vehicle types, and battery behaviors found in the global EV market.
To address these aspects, future work will focus on expanding the dataset to include a wider variety of vehicles and driving environments, enhancing the robustness of the machine learning models, and investigating potential countermeasures to reduce the privacy risks identified in this study.
Moreover, as EV security continues to advance, further research will be necessary to explore the interplay between battery data privacy and other data sources, such as charging stations, vehicle-to-vehicle communications, and in-vehicle sensors, to develop a comprehensive understanding of the security landscape in connected vehicles.
\bibliographystyle{splncs04}
\bibliography{bibliography}
\end{document}